\documentclass[pra,twocolumn,tbtags,superscriptaddress,floatfix]{revtex4}
\usepackage{amssymb}
\usepackage{graphicx,amsmath}
\usepackage{bm}
\usepackage{times}

\graphicspath{{Figures/}}
\begin{document}
\title{Influence of Impurity on the Rate of Single Photon Superradiance in Disordered
N Qubit Chain}

\begin{abstract}
We investigate the rate of superradiant emission for a number of
artificial atoms (qubits) embedded in a one-dimensional open
waveguide. More specifically, we study the 1D (N+1)- qubit chain
where N qubits are identical in respect to their excitation
frequency $\Omega$ but have different rates of spontaneous
emission $\Gamma_n$, and a single impurity qubit which is
different from N qubits by its excitation frequency  $\Omega_P$
and rate of spontaneous emission $\Gamma_P$. This system is shown
to have two hybridized collective states which accumulates the
widths of all qubits. The energy spectrum of these states and
corresponding probabilities are investigated as the function of
the frequency detuning between the impurity and other qubits in a
chain. It is shown that the inclusion of impurity qubit alter
 the resonance widths of the system only in a narrow
range of the frequency detuning between qubits and impurity, where
the resonance widths experience a significant repulsion.  The
photon transmission through disordered N- qubit chain with
impurity qubit is also considered. It is shown that a single
photon transport through this system is described by a simple
expression which predicts for specific photon frequency the
existence of a complete transmission peak and transparency window
between frequencies $\Omega$ and $\Omega_P$.


\end{abstract}


\date{\today }
\author{Ya. S. Greenberg}\email{greenbergy@risp.ru}
\affiliation{Novosibirsk State Technical University, Novosibirsk,
Russia}

\author{A. G. Moiseev}
\affiliation{Novosibirsk State Technical University, Novosibirsk,
Russia}


 \maketitle
 \section{Introduction}\label{Intr}
The phenomenon of superradiance was first discovered by R. H.
Dicke \cite{Dick54}, who showed that the system of N identical two
level excited atoms undergoes a spontaneous coherent transition to
the ground state. This is accompanied by the emission of N
photons, the intensity of which scales as $N^2$, and the decay
rate of which is $N\gamma$ , where $\gamma$ is the decay rate of
an isolated atom. In \cite{Scull09} the possibility of a single
photon superradiance was indicated, which is realized when a
single photon is sent through a qubit chain the dimension of which
is much less the photon wavelength. In this case a single photon
Dicke state is formed when N identical two level atoms are in a
symmetrical superposition of the states with one excited atom and
N-1 atoms in the ground state. In this case, the decay rate of a
single photon is also equal to $N\gamma$.

However, unlike the real atoms, qubits are intrinsically not
identical due to technological scattering of their parameters.
While the excitation energy of every qubit in a chain can, in
principle, be adjusted by external circuit, the decay rate of
every qubit is individual and cannot be controlled externally.
Nevertheless, an array of many intrinsically disordered artificial
atoms globally coupled to a single waveguide mode reveals the
existence of collective quantum behavior corresponding to coherent
oscillations of qubits \cite{Macha14, Shapiro15}.

In this paper we consider a one dimensional $N+1$ qubit chain in
an open waveguide, where $N$ qubits have the same excitation
frequency $\Omega$  and are disordered with respect to their rates
of spontaneous emission $\Gamma_n$, while a single impurity qubit
has different excitation frequency $\Omega_p$ and the rate of
spontaneous emission $\Gamma_p$. The incident photon is directed
along waveguide axis. The qubits are not coupled directly. They
interact only via a common photon field in a waveguide. The goal
of our study is to investigate the influence of impurity qubit on
the formation of superradiant state and its characteristics. The
system under study is described by a non Hermitian Hamiltonian
which accounts for the irreversible decay of the excited qibits to
the continuum.

The paper is organized as follows. In the Section II we find the
collective states for N disordered qubits in the chain. Inspite of
the fact that for N qubits all rates of spontaneous emission are
different there exists a single collective state which accumulates
all the qubits' widths. Other N-1 states are nondecaying
degenerate states with the same energy, $\hbar\Omega$. In Section
III the Hilbert space is enlarged to include the impurity qubit in
the system. The presence of the impurity qubit results in a
hybridization of its state with the superradiant collective state
of N qubit chain. We investigated the dependence of the energy
spectrum of this enlarged system on the detuning
$\Delta\Omega=\Omega-\Omega_P$. At the end of the section we
calculated the contribution of N- qubit state and impurity
wavefunction to the formation of hybridized states depending on
the frequency detuning. In Section IV the photon transmission
through disordered N- qubit chain with impurity qubit is
considered. It is shown that a single photon transport through
this system is described by a simple expression which predicts for
specific photon frequency the existence of a complete transmission
peak and transparency window between frequencies $\Omega$ and
$\Omega_P$.

\section{Collective states of disordered N- qubit chain}
As a basis set of state vectors we take the states where one qubit
is in the excited state $|e>$  and the other N-1 qubits are in the
ground state $|g>$ . Therefore, we have N vectors $|n> =
|g_1,g_2,.....g_{n-1},e_n,g_{n+1},....g_{N-1},g_N>$. The
spontaneous emission of the excited qubit results in a continuum
states $|k> = |g_1,g_2,.......g_{N-1},g_N,k>$ , where all qubits
are in the ground state and there is one photon in a waveguide.
The interaction of qubits with a photon field can be described by
a non-Hermitian Hamiltonian \footnotemark [2]:\footnotetext[2]{In
what follows we set  for convenience $\hbar=1$, so that all
energies are expressed in frequency units.}

\begin{equation}\label{H}
    H = {H_S} - iW
\end{equation}
where
\begin{equation}\label{HS}
{H_S} = \sum\limits_{n = 1}^N {\frac{1}{2}\left( {1 + \sigma
_z^{(n)}} \right){\Omega _n}}
\end{equation}

 is Hamiltonian of $N$ noninteracting qubits, and W describes the interaction of qubits
 with the photon field.
The matrix elements of Hamiltonian (\ref{H}) in the $|n>$
representation are:

\begin{equation}\label{ME}
\left\langle m \right|H\left| n \right\rangle  = {\Omega
_n}{\delta _{m,n}} - i\left\langle m \right|W\left| n
\right\rangle ;{}^{}{}^{}{}^{}1 \le m,n \le N
\end{equation}

If the distance between qubits along the direction of the photon
scattering (z- axis) is much less than the photon wavelength, the
matrix element on the right-hand side of Eq. (\ref{ME}) takes the
form \cite{Green15}:

\begin{equation}\label{W}
\left\langle m \right|W\left| n \right\rangle  = \sqrt {{\Gamma
_m}{\Gamma _n}}
\end{equation}

where $\Gamma_n$  is the rate of spontaneous photon emission from
a state where the n-th qubit is excited. Therefore, we get from
(\ref{ME}) a non-Hermitian $N\times N$ matrix, where the main
diagonal elements and off-diagonal elements are ${\Omega _n} -
i{\Gamma _n}$ and $ - i\sqrt {{\Gamma _n}{\Gamma _m}}$,
respectively. The incident photon, when absorbed, can excite any
qubit. As we do not know which of the N qubits is excited, the
wave functions of the system should be expressed as a
superposition of the state vectors $|n>$ :

\begin{equation}\label{Psi}
{\Psi _i} = \sum\limits_{n = 1}^N {{c_{n,i}}\left| n \right\rangle
}
\end{equation}
where $i=1,2,….N$.

In general, it is hard to obtain for large N the energy spectrum
of this system, however, the problem is much simplified if all
qubits have the same excitation energy: $\Omega_n=\Omega$. It is
not difficult to show that in this case the solution of the
Schr{\"o}dinger equation $H\Psi =E\Psi$ with $H$ and $\Psi$ from
Eqs. (\ref{H}) and (\ref{Psi}), respectively, has the following
properties.

 1) There is a single non stationary state with the
complex energy ${E_S} = \Omega  - i{\Gamma _S}$ where ${\Gamma _S}
= \sum\limits_{n = 1}^N {{\Gamma _n}}$. This resonance accumulates
decay widths of all qubits. The wave function of this state is as
follows:

\begin{equation}\label{PsiS}
\left| {{\Psi _S}} \right\rangle  = A\sum\limits_{n = 1}^N {\sqrt
{\frac{{{\Gamma _n}}}{{{\Gamma _1}}}} \left| n \right\rangle }
\end{equation}

where $A$ is a normalizing factor: $A = \sqrt {{\Gamma
_1}/{\Gamma_S}}$.

2) There are $N-1$ degenerate mutual orthogonal stationary states
with the energy $E=\Omega$:
\begin{equation}\label{Psim}
\left| {{\Psi _m}} \right\rangle  = \sum\nolimits_{n = 1}^N {\sqrt
{\frac{{{\Gamma _n}}}{{{\Gamma _1}}}} c_n^m\left| n \right\rangle
} {\rm{  }}\left( {m = 1,2,...N - 1} \right)
\end{equation}
where the coefficients $c_n^m$  in  (\ref{Psim}) satisfy $N-1$
conditions
\begin{equation}\label{cond}
\sum\nolimits_{n = 1}^N {{\Gamma _n}c_n^m}  = 0
\end{equation}
which ensures the orthogonality of the functions (\ref{PsiS}) and
(\ref{Psim}): $\left\langle {{\Psi _S}} \right.\left| {{\Psi _m}}
\right\rangle  = 0$.

If we assume the same rate of spontaneous emission for all qubits:
$\left\langle m \right|W\left| n \right\rangle  = \Gamma $, then
we get the known result \cite{Green15, Mois17}: there is a single
resonant state ${E_S} = \Omega  - iN\Gamma$, the wave function of
which is a symmetric coherent superposition of the state vectors
$|n>$ , where all quantities $c_{n,i}$ are the same:
\begin{equation}\label{PsiS1}
\left| {{\Psi _S}} \right\rangle  = \frac{1}{{\sqrt N
}}\sum\limits_{n = 1}^N {\left| n \right\rangle }
\end{equation}

The $N-1$ degenerate mutual orthogonal stationary states with the
energy $E=\Omega$  are given by the wave functions (\ref{Psim})
with $N-1$ conditions $\sum\nolimits_{n = 1}^N {c_n^m}  = 0$.

The collective state (\ref{PsiS1}), which we call a single photon
Dicke state \cite{Scul06}, is formed by a single photon, which
propagates through the N qubit chain. Therefore, under this
condition, the state (\ref{PsiS1}) decays with a rate  which is N
times faster than the rate of the spontaneous emission of a single
qubit.

\section{Hybridized states of disordered N- qubit chain and the impurity}
We consider N qubit chain, where all qubits have the same
excitation frequency $\Omega$  but different rates of spontaneous
emission $\Gamma_n$. The parameters of the impurity qubit we
denote as  $\Omega_p$ and $\Gamma_p$. The non Hermitian
Hamiltonian for the whole system reads:
\begin{equation}\label{H1}
H = {H_S} - iW + {H_p} - i{W_p}
\end{equation}
The first two terms in right hand side of (\ref{H1}) are just
Hamiltonian (\ref{H}) for N- qubit chain, $H_p$ is a Hamiltonian
of impurity qubit, $W_p$ describes the decay of impurity qubit via
its direct interaction with a photon field and indirectly via its
photon mediated interactions with other qubits in a chain. The
matrix elements of Eq. (\ref{H1}) in the $|n >$ representation
are:
\begin{equation}\label{ME1}
\begin{array}{l}
\left\langle m \right|H\left| n \right\rangle  = \Omega {\delta
_{m,n}} - i\left\langle m \right|W\left| n \right\rangle  = \Omega
{\delta _{m,n}} - i\sqrt {{\Gamma _n}{\Gamma _m}} \\[0.2 cm]
\left\langle p \right|H\left| n \right\rangle  =  - i\left\langle p \right|{W_p}\left| n \right\rangle  =  - i\sqrt {{\Gamma _p}{\Gamma _n}} \\[0.2 cm]
\left\langle p \right|H\left| p \right\rangle  = {\Omega _p} - i\left\langle p \right|{W_p}\left| p \right\rangle  = {\Omega _p} - i{\Gamma _p}\\
\end{array}
\end{equation}
where $|p>$ is the state of impurity qubit.

\subsection{CALCULATION OF THE ENERGY SPECTRUM}
It is convenient to write (\ref{ME1}) in explicit matrix form as a
sum of two $(N+1)\times (N+1)$ matrices $H=H_0+V$:
\begin{equation}\label{H0}
{H_0} = \left( {\begin{array}{*{20}{c}}
{{\Omega _p} - i{\Gamma _p}}&0&0& \ldots &0\\
0&{\Omega  - i{\Gamma _1}}&{ - i{\Gamma _{12}}}& \ldots &{ - i{\Gamma _{1N}}}\\
0&{ - i{\Gamma _{21}}}&{\Omega  - i{\Gamma _2}}& \ldots &{ - i{\Gamma _{2N}}}\\
 \vdots & \vdots & \vdots & \ddots & \vdots \\
0&{ - i{\Gamma _{N1}}}&{ - i{\Gamma _{N2}}}& \ldots &{\Omega  -
i{\Gamma _N}}
\end{array}} \right)
\end{equation}

\begin{equation}\label{V}
    {\rm{    }}V = \left( {\begin{array}{*{20}{c}}
0&{ - i{\Gamma _{p1}}}&{ - i{\Gamma _{p2}}}& \ldots &{ - i{\Gamma _{pN}}}\\
{ - i{\Gamma _{1p}}}&0&0& \ldots &0\\
{ - i{\Gamma _{2p}}}&0&0& \ldots &0\\
 \vdots & \vdots & \vdots & \ddots & \vdots \\
{ - i{\Gamma _{Np}}}&0& \ldots & \ldots &0
\end{array}} \right)
\end{equation}
where ${\Gamma _{mn}} = {\Gamma _{nm}} = \sqrt {{\Gamma _m}{\Gamma
_n}}$, ${\Gamma _{pn}} = {\Gamma _{np}} = \sqrt {{\Gamma
_p}{\Gamma _n}}$. The matrix $V$ exhibits photon mediated
interaction between impurity atom and other N qubits.

The basis wave functions for the system (N qubits + impurity atom)
are the states   (\ref{PsiS}) and   (\ref{Psim}), and the state of
the impurity atom, which is denoted as $\left| {{\varphi _p}}
\right\rangle$ . For subsequent calculations we express these
states as the N+1-component column vectors:
\begin{equation}\label{Psi2}
\begin{array}{l}
\left| {{\varphi _p}} \right\rangle  = \left(
{\begin{array}{*{20}{c}}
1\\
\begin{array}{l}
0\\
0
\end{array}\\
 \ldots \\
0\\
0
\end{array}} \right){\rm{;}}\left| {{\Psi _S}} \right\rangle  = \frac{A}{{\sqrt {{\Gamma _1}} }}\left( {\begin{array}{*{20}{c}}
0\\
{\sqrt {{\Gamma _1}} }\\
\begin{array}{l}
\sqrt {{\Gamma _2}} \\
{\rm{ }}....
\end{array}\\
{\sqrt {{\Gamma _{N - 1}}} }\\
{\sqrt {{\Gamma _N}} }
\end{array}} \right);\\
\,\left| {{\Psi _n}} \right\rangle  = \frac{1}{{\sqrt {{\Gamma
_1}} }}\left( {\begin{array}{*{20}{c}}
0\\
\begin{array}{l}
\sqrt {{\Gamma _1}} c_1^n\\
\sqrt {{\Gamma _2}} c_2^n
\end{array}\\
 \ldots \\
{\sqrt {{\Gamma _{N - 1}}} c_{N - 1}^n}\\
{\sqrt {{\Gamma _N}} c_N^n}
\end{array}} \right)
\end{array}
\end{equation}
It can be easily verified that the wave functions (\ref{Psi2}) are
the eigenfunctions for Hamiltonian $H_0$: ${H_0}\left| {{\varphi
_p}} \right\rangle  = ({\Omega _p} - i{\Gamma _p})\left| {{\varphi
_p}} \right\rangle$ , ${H_0}\left| {{\Psi _n}} \right\rangle  =
\Omega \left| {{\Psi _n}} \right\rangle$, ${H_0}\left| {{\Psi _S}}
\right\rangle  = \left( {\Omega  - i{\Gamma _S}} \right)\left|
{{\Psi _S}} \right\rangle$. The decomposition of $H$ into $H_0$
and $V$ is justified by the fact that due to the conditions
(\ref{cond}) the dark states $\left| {{\Psi _n}} \right\rangle$
are unaffected by the interaction term $V$. Hence, the interaction
matrix $V$ mixes up only the states $\left| {{\Psi _S}}
\right\rangle$ and $\left| {{\varphi_p}} \right\rangle$.
Therefore, we express the solution of Hamiltonian $H$ as a
hybridized state
\begin{equation}\label{hybr}
\left| \Psi  \right\rangle  = {C_p}\left| {{\varphi _p}}
\right\rangle  + {C_S}\left| {{\Psi _S}} \right\rangle
\end{equation}
The energies $E$ and the superposition factors $C_P$, $C_S$ can be
found from Schrodinger equation $H\left| \Psi  \right\rangle  =
E\left| \Psi  \right\rangle$
  for $H$ (\ref{H0}), (\ref{V}) and $\left| {\varphi _p } \right\rangle ,{\rm{ }}\left| {\Psi _S }
\right\rangle $  (\ref{Psi2}). Thus, we obtain for the
coefficients $C_P$, $C_S$ in (\ref{hybr}) two coupled equations:
\begin{equation}\label{Eqn1}
\left( {\Omega _p  - i\Gamma _p  - E} \right)C_p  - i\sqrt {\Gamma
_p \Gamma _s } \;C_S  = 0
\end{equation}
\begin{equation}\label{Eqn2}
  - i\sqrt {\Gamma _p \Gamma _s } {\rm{ }}C_p  + \left( {\Omega  - i\Gamma _s  - E} \right)C_S  = 0
\end{equation}
The energies $E$ are found by equating determinant of this system
to zero. The result is as follows:
\begin{equation}\label{Eng}
\begin{array}{l}
 E_ \pm   = \Omega  - \frac{{\Delta \Omega }}{2} - \frac{i}{2}\left( {\Gamma _S  + \Gamma _P } \right)
 \\[0.2 cm]
 {\rm{     }} \pm \frac{1}{2}\sqrt {[\Delta \Omega  - i\left( {\Gamma _S  - \Gamma _P } \right)]^2  - 4\Gamma _S \Gamma _{_P } }  \\
 \end{array}
\end{equation}
where $\Delta\Omega=\Omega-\Omega_p$.

In what follows we analyze the behavior of real and imaginary
parts of complex resonances $E_{\pm}$  of Eq. (\ref{Eng}) as a
function of the impurity detuning $\Delta\Omega$. We show that the
widths of two resonances experience the repulsion in a narrow
range of $\Delta\Omega$ near zero. The effect of repulsion of
resonance widths in open systems is known for a long time
\cite{Brent99, Volya03}. Due to this effect the impurity atom
gives rise to the enhancement of the rate of superradiance emitted
by assembly of N qubits.

First, using the method described in \cite{Brent99}, we obtain
from (\ref{Eng}):
\begin{equation}\label{DE}
\left( {E_ +   - E_ -  } \right)^2  = \left( {\Delta \Omega }
\right)^2  - \left( {\Gamma _S  + \Gamma _p } \right)^2  -
2i\Delta \Omega \left( {\Gamma _S  - \Gamma _p } \right)
\end{equation}
Second, we explicitly split $E_{\pm}$  into real and imaginary
parts
\begin{equation}\label{ReIm}
 E_ +   = \varepsilon _ +   - i\gamma _ +  ;\;E_ -   =
\varepsilon _ -   - i\gamma _ -
\end{equation}

Summing up these two equations and using explicit expression
(\ref{Eng}) for $E_{}\pm$, we obtain:
\begin{equation}\label{21}
    \varepsilon_++\varepsilon_-=\Omega-\frac{\Delta\Omega}{2};\;
    \gamma_++\gamma_-=\Gamma_S+\Gamma_P
\end{equation}
Second equation in (\ref{21}) is of general nature: total width of
all resonances has to be equal to the sum of all individual width
\cite{Sokol92}.

Then, from (\ref{DE}) we obtain two coupled equations for $\Delta
\varepsilon  = \varepsilon _ +   - \varepsilon _ - $   and $
\Delta \gamma  = \gamma _ +   - \gamma _ - $ :

\begin{equation}\label{Dre}
(\Delta \varepsilon )^2  - (\Delta \gamma )^2  = \left( {\Delta
\Omega } \right)^2  - \left( {\Gamma _S  + \Gamma _p } \right)^2
\end{equation}

\begin{equation}\label{Dim}
\Delta \varepsilon \Delta \gamma  = \Delta \Omega \left( {\Gamma
_S  - \Gamma _p } \right)
\end{equation}

We could solve these equations for $\Delta\epsilon$   and
$\Delta\gamma$ , however, we prefer to analyze their properties
and show the behavior of $\epsilon_{\pm}$ and $\gamma_{\pm}$ on
the graphs below. First, we notice from (\ref{Dim}) that at the
point $\Delta\Omega=0$, the only solution which is consistent with
(\ref{Dre}) is $\Delta\epsilon=0$,
$\Delta\gamma=\pm(\Gamma_S+\Gamma_P)$. If $\Gamma_S=\Gamma_P$
there are two possibilities: 1) $\Delta\gamma=0$,
$\Delta\epsilon\neq 0$ for $|\Delta\Omega|>2\Gamma_P$, and 2)
$\Delta\gamma\neq 0$, $\Delta\epsilon=0$ for
$|\Delta\Omega|<2\Gamma_P$.

For the case  $\Delta\Omega=0$, we obtain from (\ref{Eng}) the
expected result: $E_ +   = \Omega, E_ -   = \Omega  - i(\Gamma _S
+ \Gamma _P )$. If detuning is large ($\Delta\Omega \gg \Gamma_S)$
we obtain from (\ref{Eng}): $E_+   = \Omega - i\Gamma_S , E_- =
\Omega_P  - i\Gamma _P$. This result is also reasonable, since for
large detuning two systems (the impurity qubit and N- qubit chain
must be independent.

Below we study the dependence of the real and imaginary parts of
energies (\ref{Eng}) on the detuning $\Delta\Omega$. For above
quantities we take the values which are relevant for
superconducting qubits which operate in a microwave domain. We
take $\Omega/2\pi=5$ GHz, $\Gamma_{av}/2\pi =5$ MHz, where we
introduce the average rate $\Gamma_{av}=\Gamma_S/N$. The
dependence of real parts of the energies $E_{\pm}$ (\ref{Eng}) on
the detuning    is shown in Fig. 1 for five qubits with the same
excitation frequency   and one impurity qubit with excitation
frequency  $\Omega_p$.

\begin{figure} [h]
  \includegraphics[width=8 cm]{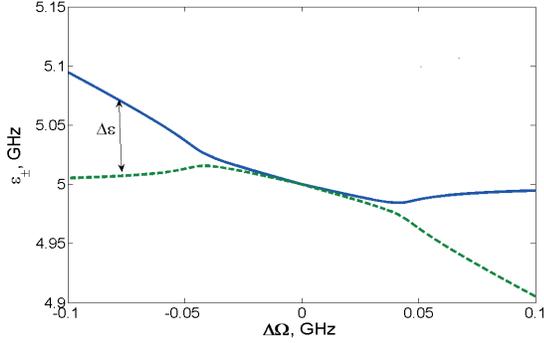}\\
  \caption{Color online. The dependence of real parts of energies $E_{\pm}$ (\ref{Eng})
   on the detuning. N=5, $\Gamma_P/\Gamma_{av}=4$,  $\Omega/2\pi=5$ GHz,
  $\Gamma_{av}/2\pi=5$MHz. Blue (solid) line is for $\epsilon_+$,
   green (dashed) line is for $\epsilon_-$}\label{f1}
\end{figure}

Two lines in Fig.\ref{f1} do not intersect but only touch each
other in a single point $\Delta\Omega=0$.

\begin{figure} [h]
   \includegraphics[width=8 cm]{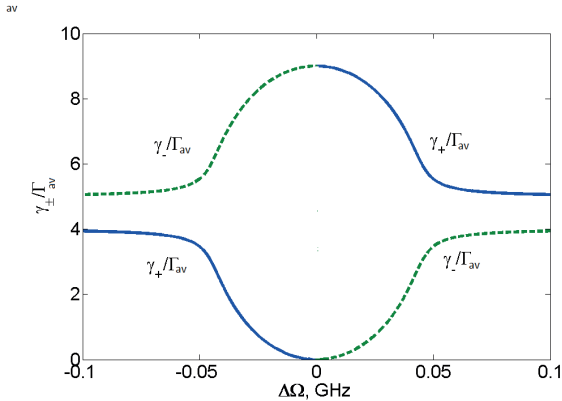}\\
   \caption{Color online. The dependence of normalized imaginary parts of
   energies, $\gamma_{\pm}/\Gamma_{av}$, on the
   detuning. N=5, $\Gamma_P/\Gamma_{av}=4$,  $\Omega/2\pi=5$ GHz,
   $\Gamma_{av}/2\pi=5$MHz. Blue (solid) line is for $\gamma_+$,
    green (dashed) line is for $\gamma_-$}\label{f2}
 \end{figure}

The dependence of normalized widths
on the detuning is shown in Fig.\ref{f2}.
   We note from this figure that the widths of two resonances experience the repulsion in a
   narrow range of $\Delta\Omega$  near zero. It is seen that the transition from the width
   $\Gamma_S+\Gamma_P$ (for the point $\Delta\Omega=0$) to the width $\Gamma_S$, and
   from zero width to $\Gamma_p$ for the impurity occurs in a narrow frequency range
    of frequency detuning not exceeding one percent from the qubit excitation
    frequency. Another interesting point is that the "phases" of two complex
    resonances (\ref{Eng}) experience a jump when crossing zero
    point of detuning. The reason for this is  that the quantity $\Delta\epsilon$
    does not change the sign when crossing the zero detuning point
    $\Delta\Omega=0$ as is seen from Fig.\ref{f1}. Then, from equation
    (\ref{Dim}) it follows that the quantity $\Delta\gamma$ must
    change its sign together with the sign of $\Delta\Omega$.

\begin{figure} [h]
     \includegraphics[width=8 cm]{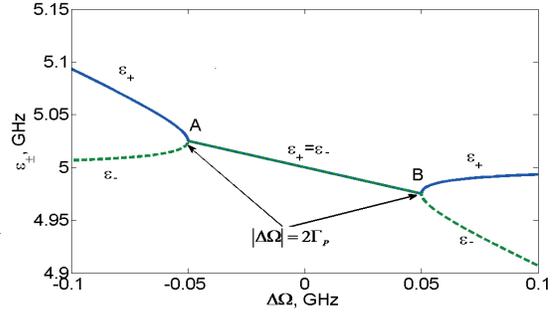}\\
     \caption{Color online. The dependence of real parts of energies on the
     detuning. N=5, $\Gamma_P/\Gamma_{av}=5$,  $\Omega/2\pi=5$ GHz,
     $\Gamma_{av}/2\pi=5$ MHz. Blue (solid) line is for $\epsilon_+$,
      green (dashed) line is for $\epsilon_-$}\label{f3}
   \end{figure}

\begin{figure} [h]
      \includegraphics[width=8 cm]{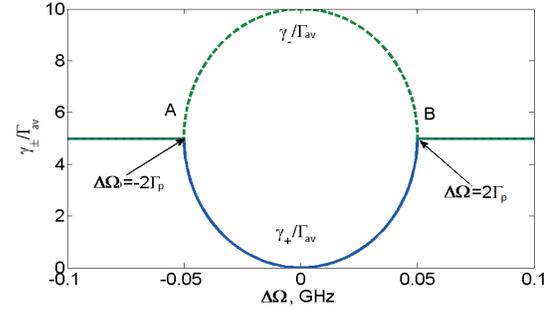}\\
      \caption{Color online. The dependence of normalized imaginary
      parts of energies, $\gamma_{\pm}/\Gamma_{av}$, on the
      detuning. N=5, $\Gamma_P/\Gamma_{av}=5$,  $\Omega/2\pi=5$ GHz,
      $\Gamma_{av}/2\pi=5$MHz. Blue (solid) line is for $\gamma_+$,
       green (dashed) line is for $\gamma_-$}\label{f4}
    \end{figure}

Below we consider the behavior of complex energies (\ref{Eng}) for
$\Gamma_S=\Gamma_P$. As is shown in Fig.\ref{f3}, the real parts
of the complex energies are merged between the points A and B
where $|\Delta\Omega|< \Gamma_p$. However, in this region the
resonance widths are repulsed reaching the maximum repulsion
$\Gamma_S+\Gamma_P$ at zero detuning  $\Delta\Omega=0$ (see
Fig.\ref{f4}). In the region where $|\Delta\Omega|> \Gamma_p$, the
picture is different: the real parts of energies are repulsed
while the widths are merged.

     \subsection{Calculation of probability factors}
Here we find the probability factors $C_P$ and $C_S$ in
(\ref{hybr}). The eigenenergies (\ref{Eng}) correspond to two
eigenfunctions

\begin{equation}\label{hybr1}
\left| {\Psi _ \pm  } \right\rangle
 = C_p^ \pm  \left| {\varphi _p } \right\rangle  + C_S^ \pm  \left| {\Psi _S }
 \right\rangle
\end{equation}

Since non- Hermitian Hamiltonian $H$  is symmetric (see (\ref{H0}), (\ref{V})),
$H = H^t$, we have $H^{\dagger} = H^{\ast}$ with the
consequence that the dual basis states of $H^{\dagger}$, $\left| {\Phi _ \pm  }
\right\rangle$  can be chosen as \cite{Keck03}:

\begin{equation}\label{Fplus}
     \left| {\Phi _ \pm  } \right\rangle
 = \left| {\Psi _ \pm ^ *  } \right\rangle  = C_p^{ \pm  * }
\left| {\varphi _p } \right\rangle  + C_S^{ \pm  * } \left| {\Psi _S } \right\rangle
\end{equation}

\begin{equation}\label{Fminus}
\left\langle {\Phi _ \pm  } \right| = \left\langle {\Psi _ \pm ^ *  }
\right| = C_p^ \pm  \left\langle {\varphi _p }
 \right| + C_S^ \pm  \left\langle {\Psi _S } \right|
\end{equation}

with the orthogonality conditions

\begin{equation}\label{ort}
\left\langle {\Psi _ \pm ^ *  |\Psi _ \pm ^{} } \right\rangle
 = 1;\quad \left\langle {\Psi _ \pm ^ *  |\Psi _ \mp ^{} } \right\rangle  = 0
\end{equation}

From (\ref{ort}) we obtain

\begin{equation}\label{norm1}
\left( {C_p^ \pm  } \right)^2  + \left( {C_S^ \pm  } \right)^2  = 1
\end{equation}

\begin{equation}\label{ort1}
    C_p^ \pm  C_p^ \mp   + C_S^ \pm  C_S^ \mp   = 0
\end{equation}

Using the equation (\ref{Eqn1}) in the form
\begin{equation}\label{29}
    \left( {\Omega _p  - i\Gamma _p  - E_ \pm  } \right)C_P^ \pm   -
i\sqrt {\Gamma _p \Gamma _s } \;C_S^ \pm   = 0
\end{equation}

and the conditions (\ref{norm1}), (\ref{ort1}), we find the
following expressions for $ C_p^ \pm  ,\;C_S^ \pm $:

\begin{equation}\label{30}
C_P^ \pm   =  \mp i\frac{{\Omega  - i\Gamma _S  - E_ \pm }}{{\sqrt
{\Gamma _S \Gamma _P  - (\Omega  - i\Gamma _S  - E_ \pm )^2 } }}
\end{equation}

\begin{equation}\label{31}
C_S^ \pm   =  \pm \frac{{\sqrt {\Gamma _S \Gamma _P } }}{{\sqrt
{\Gamma _S \Gamma _P  - (\Omega  - i\Gamma _S  - E_ \pm  )^2 } }}
\end{equation}

The orthogonality condition (\ref{ort1}) can be written as

\begin{equation}\label{32}
(\Omega  - i\Gamma _s  - E_ +  )(\Omega  - i\Gamma _s  - E_ -  ) -
\Gamma _S \Gamma _P  = 0
\end{equation}

which can be proved analytically with the aid of explicit
expression (\ref{Eng}) for $E_{\pm}$.

The equation (\ref{32}) allows one to rewrite the factors  $ C_p^
\pm  ,\;C_S^ \pm $ in a following form:

\begin{equation}\label{33}
C_P^ \pm   =  \mp i\frac{{\sqrt {\Omega  - i\Gamma _S  - E_ \pm  }
}}{{\sqrt { \pm (\Delta \varepsilon  - i\Delta \gamma )} }}
\end{equation}

\begin{equation}\label{34}
C_S^ \pm   =  \pm \frac{1}{{\sqrt { \pm (\Delta \varepsilon  -
i\Delta \gamma )} }}\frac{{\sqrt {\Gamma _P \Gamma _S } }}{{\sqrt
{\Omega  - i\Gamma _S  - E_ \pm  } }}
\end{equation}

where $\Delta\epsilon$   and $\Delta\gamma$   are defined in
(\ref{ReIm}).

As is seen from (\ref{33}) and (\ref{34}) the quantities $ \left|
{C_P^ \pm  } \right|^2 ,\,\left| {C_S^ \pm  } \right|^2 $ cannot
be taken as a measure of the probabilities, since they can be
greater than one, especially for the case $\Gamma_S=\Gamma_P$  in
the vicinity of exceptional points $\Delta \Omega  =  \pm 2\Gamma
_P $ , where $\Delta \varepsilon  = \Delta \gamma  = 0$. The
reason for this is that the wave function (\ref{hybr1}) of non
Hermitian Hamiltonian (\ref{H0}), (\ref{V}) cannot be normalized
to unity \cite{Brody14}. Therefore, we define the corresponding
probabilities as follows:

\begin{equation}\label{35}
P_P^ \pm   = \frac{{\left| {\left\langle {\varphi _P |\Psi _ \pm }
\right\rangle } \right|^2 }}{{\left| {\left\langle {\Psi _ \pm
|\Psi _ \pm  } \right\rangle } \right|^2 }} = \frac{{\left| {C_P^
\pm  } \right|^2 }}{{\left| {C_P^ \pm  } \right|^2  + \left| {C_S^
\pm  } \right|^2 }}
\end{equation}

\begin{equation}\label{36}
P_S^ \pm   = \frac{{\left| {\left\langle {\Psi _S |\Psi _ \pm  }
\right\rangle } \right|^2 }}{{\left| {\left\langle {\Psi _ \pm
|\Psi _ \pm  } \right\rangle } \right|^2 }} = \frac{{\left| {C_S^
\pm  } \right|^2 }}{{\left| {C_P^ \pm  } \right|^2  + \left| {C_S^
\pm  } \right|^2 }}
\end{equation}

The explicit forms for these probabilities read:

\begin{equation}\label{37}
P_P^ \pm   = \frac{{(\Omega  - i\Gamma _S  - E_ \pm  )(\Omega  +
i\Gamma _S  - E_ \pm ^ *  )}}{{(\Omega  - i\Gamma _S  - E_ \pm
)(\Omega  + i\Gamma _S  - E_ \pm ^ *  ) + \Gamma _S \Gamma _P }}
\end{equation}

\begin{equation}\label{38}
P_S^ \pm   = \frac{{\Gamma _S \Gamma _P }}{{(\Omega  - i\Gamma _S
- E_ \pm  )(\Omega  + i\Gamma _S  - E_ \pm ^ *  ) + \Gamma _S
\Gamma _P }}
\end{equation}

\begin{figure} [h]
      \includegraphics[width=8 cm]{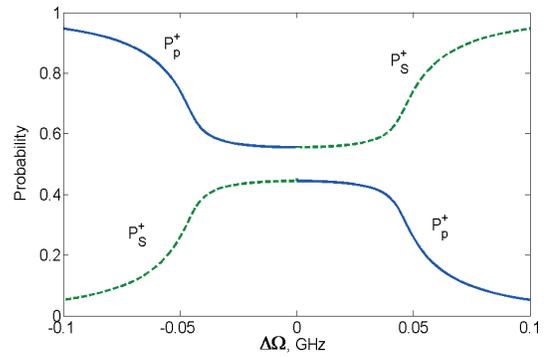}\\
      \caption{Color online. The dependence of probabilities $P_P^ + ,\,\;P_S^
      +$ for the state $|\Psi_+\rangle$
      on the detuning. N=5, $\Gamma_P/\Gamma_{av}=4$,  $\Omega/2\pi=5$ GHz,
      $\Gamma_{av}/2\pi=5$MHz. Blue (solid) line is for $P_P^+$,
       green (dashed) line is for $P_S^+$}\label{f5}
    \end{figure}

The detuning dependence of the probabilities $P_P^ + $ and $P_S^
+$ for hybridized state $|\Psi_+\rangle$ (\ref{hybr1}) is shown in
Fig.\ref{f5}. These plots were drawn for the same parameters as
those in Fig.\ref{f1} and Fig.\ref{f2}. We see that for relatively
large detuning the contribution of the one of two wave functions
($|\phi_P\rangle$ or $|\Psi_S\rangle$) to the formation of
hybridized state $|\Psi_+\rangle$ becomes dominating. For positive
detuning the contribution of the state $|\Psi_S\rangle$ dominates
the formation of hybridized state, while for negative detuning the
domination is observed for the state $|\phi_P\rangle$. For small
detuning ($\Delta\Omega<(\Gamma_S+\Gamma_P)$) the probabilities
$P_P^ + $ and $P_S^+$ are weakly dependent on detuning. At zero
detuning we observe the jump between $P_P^ + $ and $P_S^+$, the
origin of which is related to the phase jump shown in Fig.
\ref{f2}.

\begin{figure} [h]
      \includegraphics[width=8 cm]{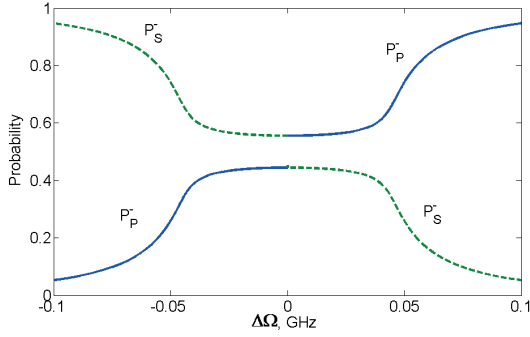}\\
      \caption{Color online. The dependence of probabilities $P_P^ - ,\,\;P_S^ -$
      for the state $|\Psi_-\rangle$ on the detuning. N=5, $\Gamma_P/\Gamma_{av}=4$,  $\Omega/2\pi=5$ GHz,
      $\Gamma_{av}/2\pi=5$MHz. Blue (solid) line is for $P_P^-$,
       green (dashed) line is for $P_S^-$}\label{f6}
    \end{figure}

The dependence of the probabilities $P_P^-$ and $P_S^-$ on the
detuning for hybridized state $|\Psi_-\rangle$ (\ref{hybr1}) is
shown in Fig. \ref{f6}. The plots were drawn for the same
parameters as those in Fig.\ref{f1} and Fig.\ref{f2}. This picture
is a mirror image of the one shown in Fig. \ref{f5}. Hence, all
properties of the plots in Fig. \ref{f6} regarding to the
contribution of the states ($|\phi_P\rangle$ and $|\Psi_S\rangle$)
to the formation of hybridized state $|\Psi_-\rangle$ are the same
(with account for mirror reflection) as those shown in Fig.
\ref{f5}.

\begin{figure} [h]
      \includegraphics[width=8 cm]{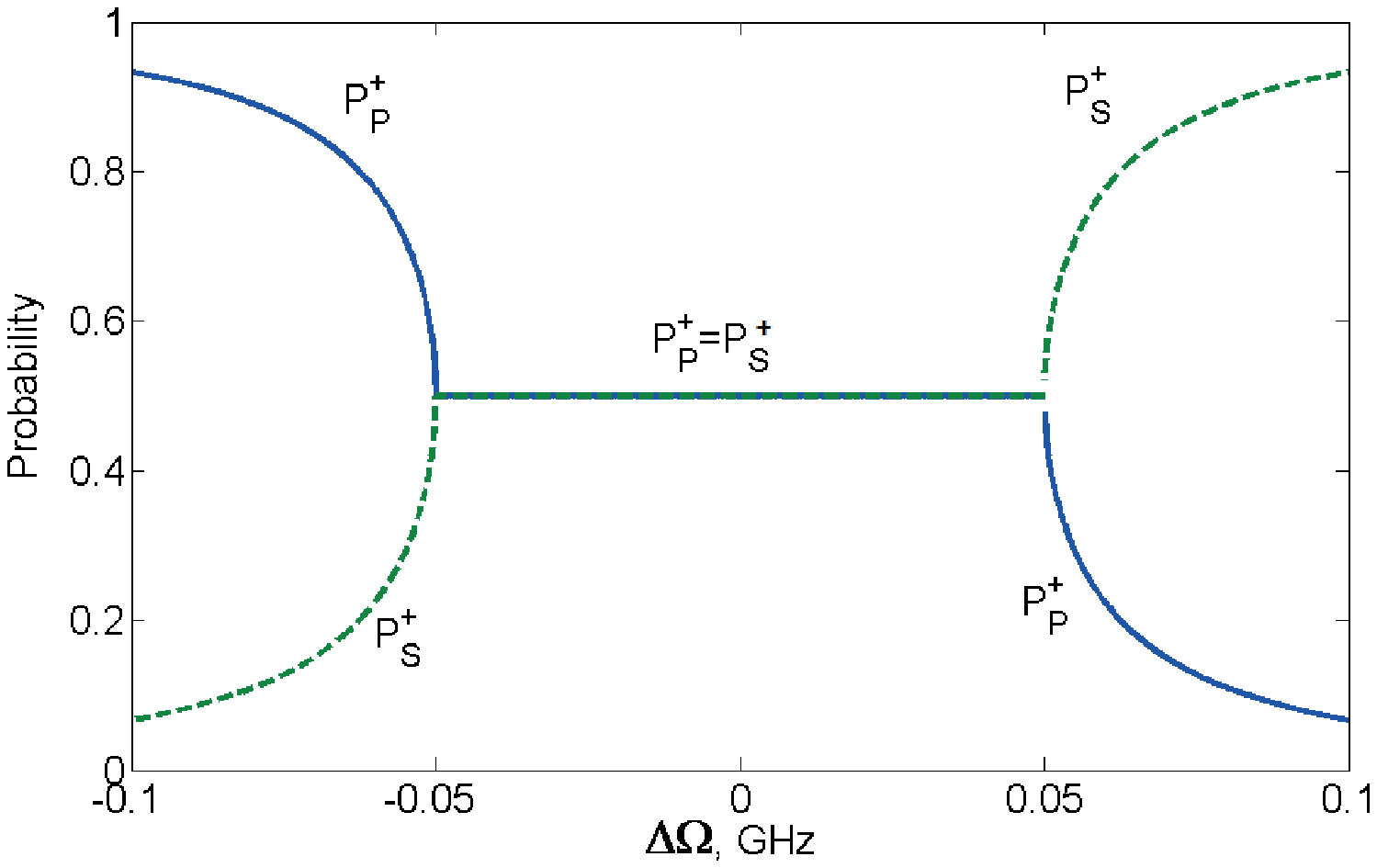}\\
      \caption{Color online. The dependence of probabilities $P_P^ + ,\,\;P_S^ +$
      on the detuning. N=5, $\Gamma_P/\Gamma_{av}=5$,  $\Omega/2\pi=5$ GHz,
      $\Gamma_{av}/2\pi=5$MHz. Blue (solid) line is for $P_P^+$,
       green (dashed) line is for $P_S^+$}\label{f7}
    \end{figure}

\begin{figure} [h]
      \includegraphics[width=8 cm]{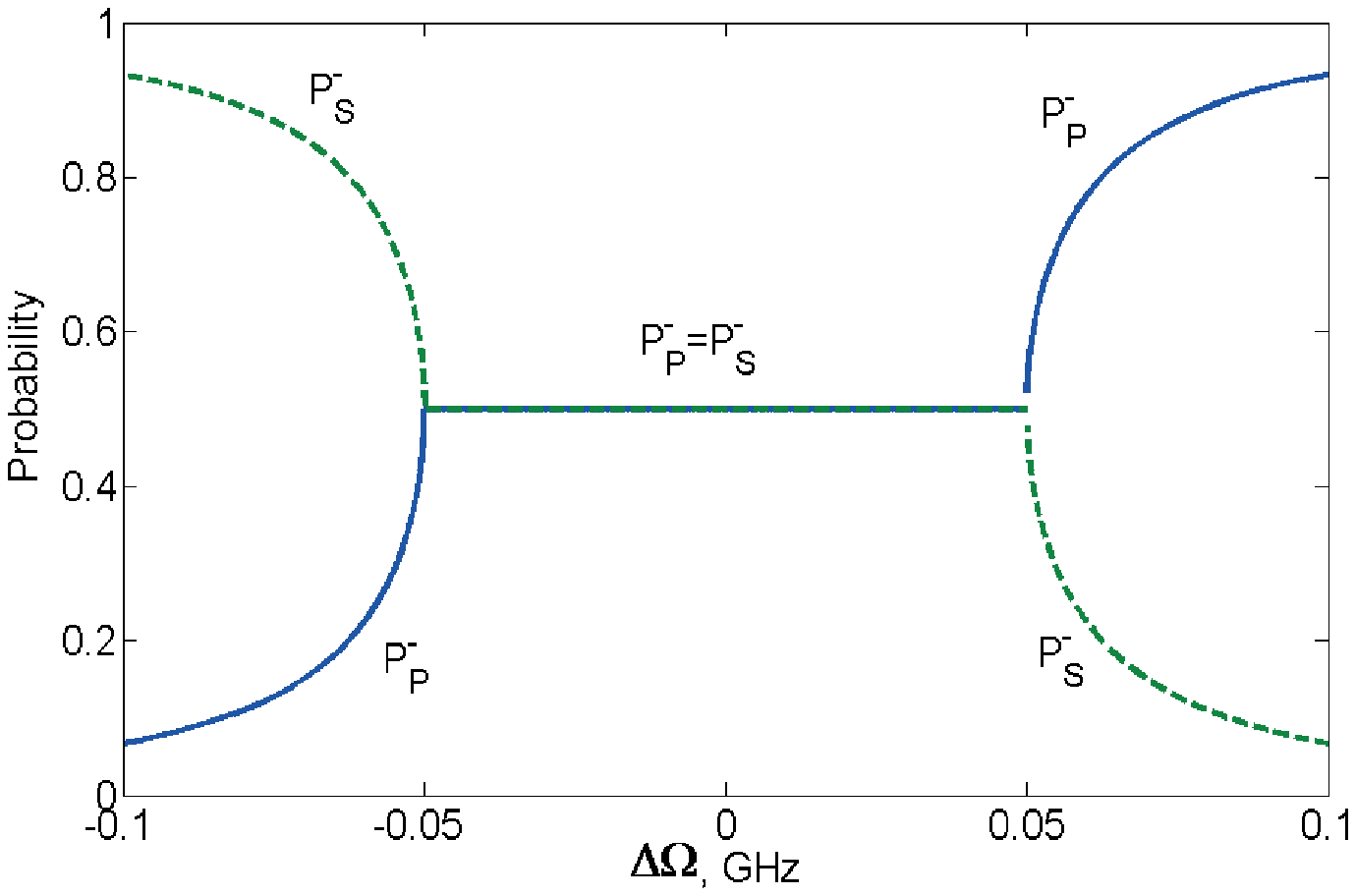}\\
      \caption{Color online. The dependence of probabilities $P_P^ - ,\,\;P_S^ -$
      on the detuning. N=5, $\Gamma_P/\Gamma_{av}=5$,  $\Omega/2\pi=5$ GHz,
      $\Gamma_{av}/2\pi=5$MHz. Blue (solid) line is for $P_P^-$,
       green (dashed) line is for $P_S^-$}\label{f8}.
    \end{figure}

Figs. \ref{f7}, \ref{f8} show the same dependences as those from
Figs. \ref{f5}, \ref{f6}, but for the case
$\Gamma_S=\Gamma_P\equiv\Gamma$. The plots of real and imaginary
parts of the complex energy for this case are shown in Figs.
\ref{f3}, \ref{f4}. We see that in the central region of these
plots, $|\Delta\Omega|<2\Gamma$, the contributions of the wave
functions  $|\phi_P\rangle$ and $|\Psi_S\rangle$ to the formation
of hybridized states $|\Psi_{\pm}\rangle$ are equal ($P_S^+=P_P^+$
in Fig. \ref{f7} and $P_S^-=P_P^-$ in Fig. \ref{f8}). It can be
shown explicitly, that for this case in the detuning range  $
\left| {\Delta \Omega } \right| \le 2\Gamma $ the following
relation holds
\begin{equation}\label{nnbm}
(\Omega  - i\Gamma  - E_ \pm  )(\Omega  + i\Gamma  - E_ \pm ^ *  )
= \Gamma ^2\nonumber
\end{equation}
Therefore, from (\ref{37}), (\ref{38}) we obtain  $ P_P^ \pm =
P_S^ \pm = 0.5 $ in this range.

    \section{Photon transport through a disordered N-qubit chain}

There are many papers where the one- photon transmission and
reflection spectra are studied for two or more qubits interacting
with waveguide, plasmonic or resonator modes. It was reported that
for some specific qubit arrays it is possible to form a reflection
or transmission window in a broad wave range \cite{Kim10, Kim14,
Chang11, Cheng17}.

Here we show that for system we study, namely, $N$ qubits with
equal excitation frequency $\Omega$ and different rates of
spontaneous emission $\Gamma_n$, and a single impurity qubit with
excitation frequency $\Omega_P$ and rate of spontaneous emission
$\Gamma_P$, the expressions for transmission and reflection
spectra are greatly simplified: they are reduced to those for two
qubits with different excitation frequencies.

In the single- photon approximation and long wavelength limit
($\omega L/v_g\ll 1$, where $L$ is the chain length, $v_g$ group
velocity of the waveguide mode) the general expression for the
transmission coefficient for N- qubit chain is as follows
\cite{Green15}:


%

    \begin{equation}\label{tr}
t_N  = \frac{{\prod\nolimits_{n = 1}^N {(\omega  - \Omega _n )}
}}{{\prod\nolimits_{n = 1}^N {(\omega  - z_n )} }}
\end{equation}
where $\omega$ is the frequency of incident photon, $z_n$ are the
eigenenergies of non Hermitian Hamiltonian (\ref{H}). For the
system studied in this paper the expression (\ref{tr}) can be
reduced to the form below:

\begin{equation}\label{tr1}
t_N  = \frac{{(\omega  - \Omega )(\omega  - \Omega _P )}}{{(\omega
- E_ +  )(\omega  - E_ -  )}}
\end{equation}
where the quantities $E_{\pm}$ are given in (\ref{Eng}).

Explicit expressions for $E_{\pm}$ allow to rewrite (\ref{tr1}) in
the following form

\begin{equation}\label{tr2}
t_N  = \frac{{x(x + \Delta \Omega )}}{{x[x + i(\Gamma _S  + \Gamma
_P )] + \Delta \Omega [x + i\Gamma _S ]}}
\end{equation}

where $x = \omega  - \Omega ;\;\Delta \Omega  = \Omega  - \Omega
_P ;\;\Gamma _S  = \sum\nolimits_{n = 1}^N {\Gamma _n }$.

Similar calculations give the expression for  reflection
amplitude:

\begin{equation}\label{Refl}
r_N  = -i\frac{{x(\Gamma _S  + \Gamma _P ) + \Delta \Omega \Gamma
_S }}{{x[x + i(\Gamma _S  + \Gamma _P )] + \Delta \Omega [x +
i\Gamma _S ]}}
\end{equation}

If $\Delta\Omega=0$, the transmission (\ref{tr2}) takes the form:

\begin{equation}\label{tr3}
t_N  = \frac{{\omega  - \Omega }}{{[\omega  - \Omega  + i(\Gamma
_S  + \Gamma _P )]}}
\end{equation}

It follows from (\ref{tr3}) that for N qubits identical in their
excitation frequencies the width of the resonance dip accumulates
the widths of all resonances in the system. It corresponds to the
point $\Delta\Omega=0$ at the plots shown in Figs. \ref{f2},
\ref{f4}, where the resulting width of the resonance at this point
becomes a sum of the widths of all N qubits, $\Gamma_S$ and the
width of the impurity, $\Gamma_P$. As it follows from (\ref{tr3})
the best way to reveal the existence of collectivized qubit state
is to measure a phase of transmission signal the magnitude of
which scales as the number of qubits which take part in the
formation of this collective state \cite{Macha14, Shapiro15}.

The expression (\ref{tr2}) describes, in fact, two qubits with the
parameters $\Omega, \Gamma_S$ and $\Omega_P, \Gamma_P$,
respectively. The interesting feature of this expression is that
there exists the photon frequency at which a complete transmission
is observed. This frequency corresponds to the point where the
imaginary part of denominator in (\ref{tr2}) (or nominator in
(\ref{Refl})) becomes zero:
\[x({\Gamma _S} + {\Gamma _P}) + \Delta \Omega {\Gamma _S} = 0\]
Therefore, we see that at the point
\begin{equation}\label{44}
{\omega_c} =\Omega  -
\Delta\Omega\frac{\Gamma_S}{\Gamma_S+\Gamma_P}
=\frac{\Omega\Gamma_P+\Omega_P\Gamma_S}{\Gamma_S+\Gamma_P}
\end{equation}
the transmission $t_N$=1. The frequency $\omega_c$ lies just in
between two frequencies , $\Omega$ and $\Omega_P$.

This property was earlier reported in \cite{Green15} for three
qubits (Fig.14 in \cite{Green15}). As was noted there the
frequencies at which complete transmission is observed, correspond
to the frequencies which provide zero to the reflection amplitude.
They do not, in general, coincide with the frequencies of the
resonances (real parts of the complex energies) found from the
system Hamiltonian.

For our system it can be shown explicitly that $\omega_C$ can be
expressed in terms of real and imaginary parts of complex energies
$E_{\pm}$ as follows:

\begin{equation}\label{omc}
{\omega _c} = \frac{{{\varepsilon _ + }{\gamma _ - } +
{\varepsilon _ - }{\gamma _ + }}}{{{\gamma _ + } + {\gamma _ - }}}
\end{equation}
The equivalence of right hand sides of Eqs. (\ref{44}) and
(\ref{omc}) can be seen if in (\ref{omc}) we express
$\varepsilon_{\pm}$ and $\gamma_{\pm}$ in terms of
$\Delta\epsilon$ and $\Delta\gamma$ and use the equation
(\ref{Dim}).

\begin{figure}
  \includegraphics[width=8 cm]{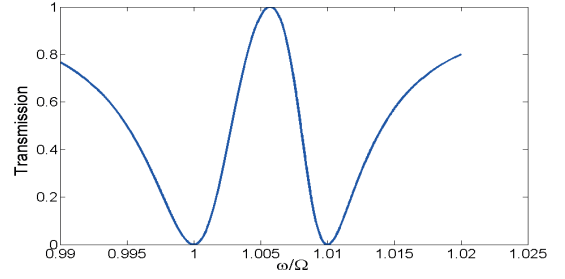}\\
  \caption{Transmission spectrum $|t|^2$ as a function of the photon frequency.
  $\Omega/2\pi=5$ GHz, $\Omega_P/2\pi=5.05$ GHz, $\Gamma_S/2\pi=20$ MHZ,
  $\Gamma_P/2\pi=15$ MHz.}\label{f9}
\end{figure}

\begin{figure}
  \includegraphics[width=8 cm]{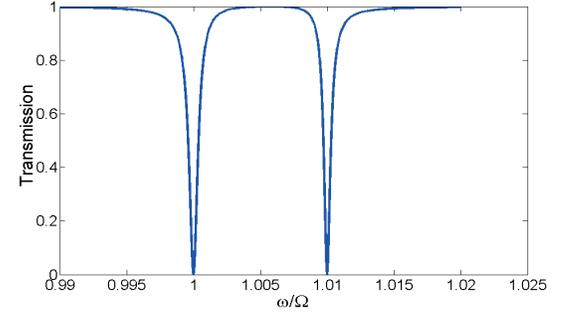}\\
  \caption{Transmission spectrum $|t|^2$ as a function of the photon frequency.
  $\Omega/2\pi=5$ GHz, $\Omega_P/2\pi=5.05$ GHz, $\Gamma_S/2\pi=2$ MHZ,
  $\Gamma_P/2\pi=1.5$ MHz.}\label{f10}
\end{figure}

The form of the transmission plot depends mainly on the relation
between two quantities $\Gamma_S (\Gamma_P)$ and the frequency
detuning $\Delta\Omega$. As the spontaneous rates $\Gamma_{S,P}$
are directly proportional to the qubit -photon interaction, we
expect that for $\Gamma_{S,P}\approx\Delta\Omega$ the plot will
have the form of interference pattern as shown in Fig.\ref{f9},
where $\Delta\Omega=50$ MHz, $\Gamma_S/2\pi=20$ MHZ,
$\Gamma_P/2\pi=15$ MHz. In the opposite case,
$\Gamma_{S,P}\ll\Delta\Omega$, we obtain between the frequencies
$\Omega$ and $\Omega_P$ a transparency window with steep walls as
shown in Fig.\ref{f10} where $\Delta\Omega=50$ MHz,
$\Gamma_S/2\pi=2$ MHZ, $\Gamma_P/2\pi=1.5$ MHz.

The formation of transparency window here is more or less
understandable, and was reported for similar structures
\cite{Kim10, Kim14}. But what is not trivial is that for our
system there always exist between the frequencies $\Omega$ and
$\Omega_P$ the frequency $\omega_c$ (\ref{44}) where the complete
transmission is observed.

\section{Conclusion}
We considered 1D N qubit chain in an open waveguide. All qubits
have equal excitation frequencies but disordered in respect to
their rates of spontaneous emission. We showed that this system
has a unique superradiant state with the width being the sum of
all individual qubits' widths. Other $N-1$ states are dark in that
they have not any widths at all and, hence, these states are not
observable. The inclusion of impurity qubit with different
excitation frequency and the rate of spontaneous emission results
in the formation of hybridized states which alter significantly
the resonance widths of the system only in a narrow range of the
frequency detuning between qubits and impurity. We also calculated
the contribution of N- qubit state and impurity wavefunction to
the formation of hybridized states depending on the frequency
detuning. We showed that a single photon transport through our
system is described by a simple expression which predicts for
specific photon frequency the existence of a complete transmission
peak and transparency window between two frequencies $\Omega$ and
$\Omega_P$.

\begin{acknowledgments}
Ya. S. G. acknowledges A. N. Sultanov for fruitful discussions.
The work is supported by Ministry of Education and Science of the
Russian Federation under Project No. 419 3.4571.2017/6.7.
\end{acknowledgments}

\end{document}